\documentclass[epjCONF,columns]{svjour} 
\usepackage{graphics}
\usepackage[varg]{txfonts} 
\usepackage[latin1]{inputenc}
\session-title{Tidal Disruption events and AGN outbursts}
\begin{document}
\title{Stability of black hole accretion disks}
\author{Agnieszka Janiuk\inst{1}\fnmsep\thanks{\email{agnes@cft.edu.pl}} \and Ranjeev Misra\inst{2} \and Bozena Czerny\inst{3} \and Magdalena Kunert-Bajraszewska\inst{4} }
\institute{Center for Theoretical Physics, Polish Academy of Sciences, Warsaw, Poland \and Inter-University Center for Astronomy and Astrophysics, Pune, India \and Nicolaus Copernicus Astronomical Center, Warsaw, Poland \and Nicolaus Copernicus University, Torun, Poland}
\abstract{
We discuss the issues of stability of accretion disks that may undergo the 
limit-cycle oscillations due to the two main types of thermal-viscous 
instabilities.
These are induced either by the domination of radiation pressure 
in the innermost regions close to the central black hole, or by 
the partial ionization of hydrogen in the zone of appropriate temperatures.
These physical processes may lead to the intermittent activity in AGN on 
timescales between hundreds and millions of years. We list a number of 
observational facts that support the idea of the cyclic activity in 
high accretion rate sources. We conclude 
however that the observed features of quasars
may provide only indirect signatures of the underlying instabilities.
Also, the support from the sources with stellar mass black holes, 
whose variability timescales are observationally feasible, is 
limited to a few cases of the microquasars. Therefore we consider a number 
of plausible mechanisms of stabilization of the limit cycle oscillations in 
high accretion rate accretion disks. The newly found is the 
stabilizing effect of 
the stochastic viscosity fluctuations. 
} 
\maketitle
\section{Introduction}
\label{intro}

The evolution and significant morphological changes in active 
galactic nuclei and their host galaxies may result from catastrophic events, 
such as mergers with other galaxies or tidal disruptions. Another
type of processes that trigger the evolution of AGN are connected with
intrinsic properties of their nuclei and do not have to involve
the companion interaction or any violent catastrophes.
In particular, the evolution of black hole accretion disk, believed
to be a power house of quasars and AGN, may lead to secular changes in the
observed properties of the source.

In this proceeding, we discuss the physics of processes in the accretion disk
around compact star that under certain conditions lead to the
thermal-viscous instability and periodic outbursts of luminosity
in the AGN central engine. We also present observational support for 
the theoretical models of the disk instability as well as we discuss some 
possible mechanisms that may act on stabilizing the system.
Because the physics is the same
for a broad range of masses of the accreting black holes, we
include in this discussion the results obtained for stellar mass black 
holes in binary systems. They are much easier to observe during the lifetime
of our currently available instruments.

\section{AGN outbursts induced by accretion instabilities}
\label{sec:1}

The outbursts of accretion disks may be induced by the two main types of 
instabilities which lead to the thermal-viscous oscillations. These are 
the radiation pressure instability and the partial hydrogen ionization
instability, both known for over 40 years in theoretical astrophysics. 
For a detailed discussion of these instabilities the reader is referred to 
the recent article \cite{Ref6} as well as to the literature quoted therein.
Below, we only briefly describe the fundamental physics behind these
processes.

\subsection{Radiation pressure instability: basic physics}
\label{sec:2}

The stationary, thin accretion disk model in classical theory is based on
$\alpha$ prescription for the viscous energy dissipation.
In the $\alpha$-model we assume the non-zero component $T_{r \phi}$
of the stress tensor is proportional to the total pressure.
The latter includes the radiation pressure component which scales 
with temperature as $T^{4}$ and blows up in hot disks for large accretion rates.
This in turn affects the heating and cooling balance, between the energy 
dissipation and radiative losses.

\begin{figure}
\resizebox{0.95\columnwidth}{!}{
 \includegraphics{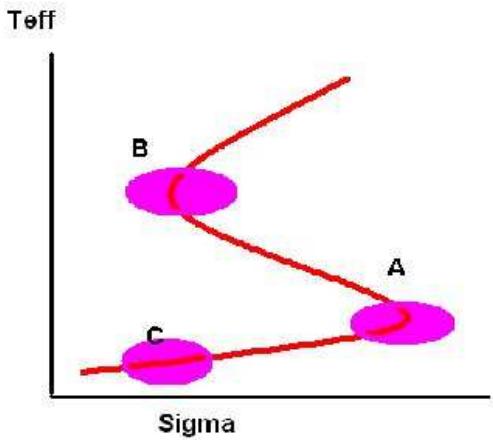}}
\caption{Schematic solutions of the local thermal balance in the accretion disk, shown on the surface density-temperature plane}
\label{fig:1}       
\end{figure}

Such a balance, under the assumption of hydrostatic equilibrium, is
calculated numerically with a closing equation for the
locally dissipated flux of energy given by the black hole mass and 
global accretion rate. The local solutions may be conveniently plotted 
on the so-called stability 
curve, shown in Figure \ref{fig:1}. The scheme consists of 
distinct points which represent the 
annulus in an accretion disk
 with temperature and surface density determined by the
accretion rate. If the accretion rate is small, i.e. induced 
temperature is  between the points A and C on this scheme, the local 
annulus of the disk is gas pressure dominated and stable. For larger 
accretion rates, and temperatures between points A and B, the annulus
will be dominated by radiation pressure and unstable.
The larger the global accretion rate, the more annuli of the disk will be
affected by the instability and the extension of the instability zone grows 
in radius, starting around the inner edge of the disk which is the hottest.

If there was no stabilizing mechanism, the radiation pressure dominated disk would not survive. This is because in such parts of the disk the 
decreasing density leads to the temperature growth. 
In consequence, the local accretion rate increases and more material is 
transported inwards. The disk annulus empties because of both increasing 
accretion rate and decreasing density, so there is no self-regulation 
of the disk structure.
However, the so called 'slim-disk' solution, where advection of energy
provides additional source of cooling in the highest accretion rate regime
(close to the Eddington limit), acts as a stabilizing branch. 
Therefore, 
even if a large part of the disk is dominated by radiation pressure, 
advection of some part of energy allows the disk to survive and oscillate 
between the hot and cold states. Such oscillating behavior leads to periodic 
changes of the disk luminosity, as shown e.g. in \cite{Ref5} for the 
black hole X-ray binary disk.
The advective solutions constitute the upper stable sequence
on the $S$-curve shown in Figure \ref{fig:1}.

\subsection{Partial hydrogen ionization instability}
\label{sec:2}

The ionization instability operates in the outer regions of the 
accretion disk, where the temperatures are in the range 
of $\log T = 3.5-4$ [K]. This is the temperature range where Hydrogen is 
partially ionized. As known from the stellar astrophysics and 
dwarf novae theory, under such conditions the opacities in the plasma
depend inversely on density and temperature, which in turn affects the local 
thermal balance, somewhat similarly to the radiation pressure discussed above.

The crucial difference between stellar black hole disks and AGN disks is
that the latter are cooler, as their global properties scale with the 
black hole mass. The AGN disks, which radiate mostly in the UV range, have 
therefore the hydrogen ionization zone much closer to the black hole than the
BHXB disks (for the distance expressed in Schwarz-schild radii). In the latter, the ionization instability zone may never form, if the disk size, 
determined by the binary system separation and orbital period, is too small. 
It may happen that the whole disk is too hot for that \cite{Ref6}.

\subsection{Overlapping of the instability zones}
\label{sec:3}

The limitation by the disk size does not apply to
 the radiation pressure instability. In consequence, at 
high accretion rates, the AGN disks may have 
two instability zones
close to each other, while in GBH there may be only the radiation pressure 
dominated unstable zone. On the other hand, 
for small accretion rates, the AGN disks will still
 have the ionization instability zone, while the GBH disks may be stable to 
whatever type of instability if they are small in radius.

\section{Observational facts}

The studies of AGN evolution are difficult with respect to any single object 
and plotting a long-term lightcurve fit to the instability model is not possible. For a typical supermassive 
 black hole of $10^{8} M_{\odot}$, 
this would require timescales of hundreds of years in case of the radiation 
pressure and millions of years for the ionization instability driven evolution.

Nevertheless, statistical studies may shed some light on the sources evolution.
For instance, the Giga-Hertz Peaked quasars, \cite{Ref3}, have very compact 
sizes which would directly imply their ages on the order of 100-1000 years. 
However, there are too many of such sources to 
be all very young. Statistically, we would have too many young sources as 
compared to the mature quasars.
An alternative explanation is therefore that in fact these sources are not 
young, but we observe now a subsequent step of the evolution, i.e. a 
reactivated source. The sample studied in \cite{Ref2} consists of 70 sources. 
The activity duty cycles are for most of them consistent with the 
sources ages, determined with a rough accuracy from either 
kinematics or synchrotron spectra.

Another possibility to study the duty cycles of quasars is studying their 
morphology and searching for various kinds of distortions or discontinuities in 
the radio structures. These structures may reflect the history of the central 
power source of a quasar, which went through the subsequent phases of activity 
and quiescence. An exemplary source of that kind, quasar 1641+320,
 was studied in \cite{Ref8} and found to exhibit multiple radio structures. 
The source image is shown in Figure \ref{fig:1}, 
where the radio structures labeled by C, E, W1 and W2 are the radio core,
the eastern and the two western radio lobes, respectively. We suggest that the 
most recent jet direction is that from C to W2 structures in this source, 
while the other lobes originated from the past activity episodes.

\begin{figure}
\resizebox{0.95\columnwidth}{!}{
 \includegraphics{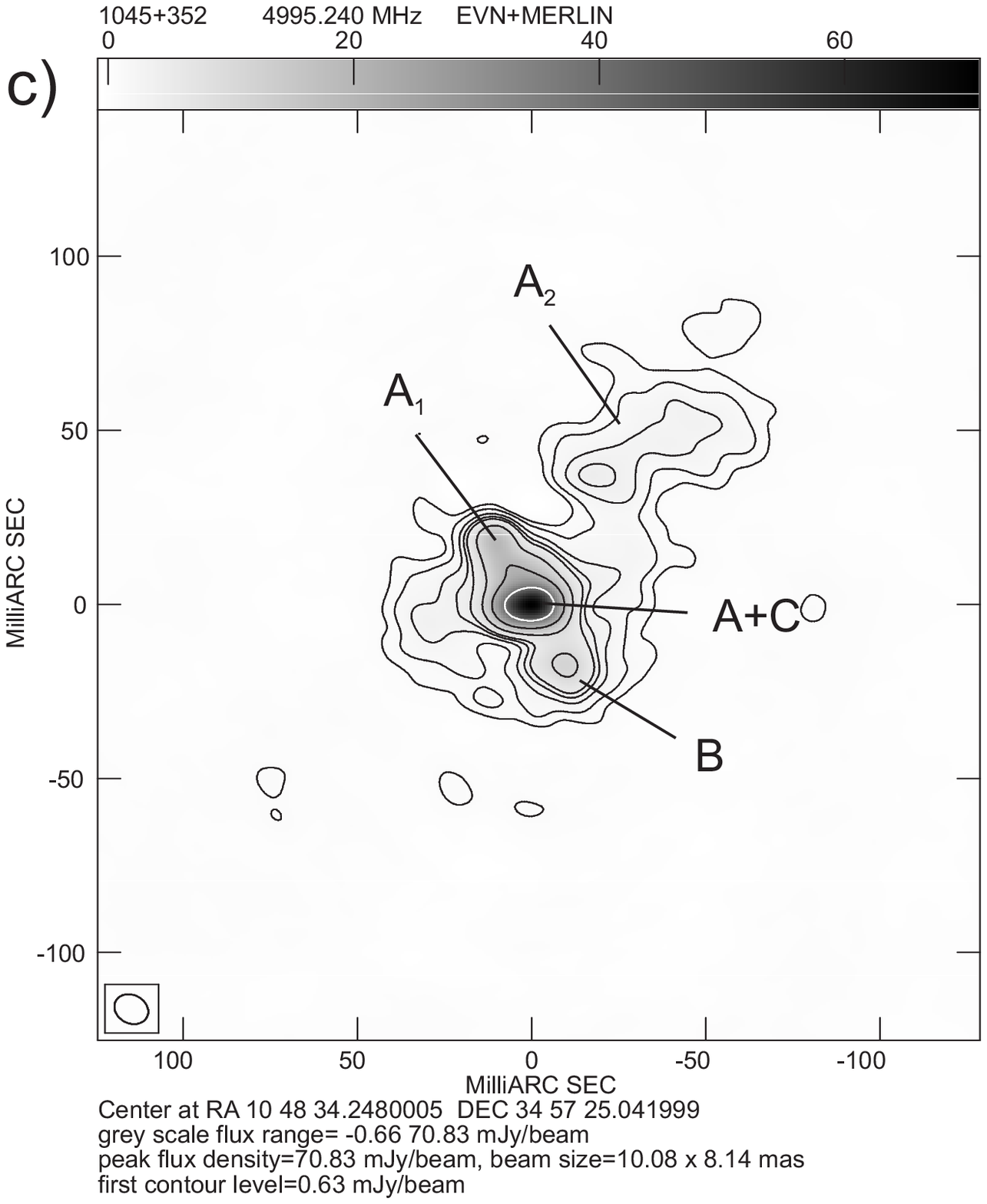}}
\caption{EVN radio image at 5 GHz of the quasar 1045+352 \cite{Ref8a}.
Contours increase by a factor of two and the first contour level
corresponds to $\sim 3\sigma$. The symbols indicate: C - radio core, A -
A2 - jet, B - counter-jet}
\label{fig:2a}       
\end{figure}

An intermittent activity scenario can explain also
the complex morphology of broad absorption line (BAL)
quasar 1045+352 \cite{Ref8a}.
The radio structure of 1045+352 is dominated by the strong radio jet 
resolved into many sub-components and changing the orientation during 
propagation in the central regions of the host galaxy. As a consequence,
 we observe at least three phases of jet activity indicate different 
directions of the jet outflow. 
The source image is shown in Figure \ref{fig:2a}, where
the radio structures labelled by C, A, $A_{1}$, $A_{2}$ and B are the
radio core and the parts of the jet, respectively. 
We suggest that the current activity direction is
the jet A emerging from the core to the east, while the other
components probably originated from the past activity episodes.

Still, the best studied example of the radiation pressure instability in action 
is the microquasar GRS 1915+105, which in some spectral states exhibits 
cyclic outbursts of X-ray luminosity, well fitted to a limit cycle oscillations 
of an accretion disk. This source is known for 20 years now, and only recently 
yet another microquasar of that type was discovered \cite{Ref1}.

All above, and some other observational hints, discussed also in more detail 
in \cite{Ref6}, support the existence of the thermal and viscous instabilities 
of accretion disks.
Nevertheless, there exist many sources accreting at high accretion rate to the 
Eddington rate ratios, that do not exhibit the limit cycle oscillations
on timescales adequate for the radiation pressure instability.
Also, many AGN do not seem to exhibit cyclic activity. This may of course 
be a selection effect, or poor statistics in the observed lightcurves.
On the other hand, some stabilizing mechanisms to the above discussed instabilities should be considered to explain the apparent stability of the high 
accretion rate sources.

\begin{figure}
\resizebox{0.95\columnwidth}{!}{
 \includegraphics{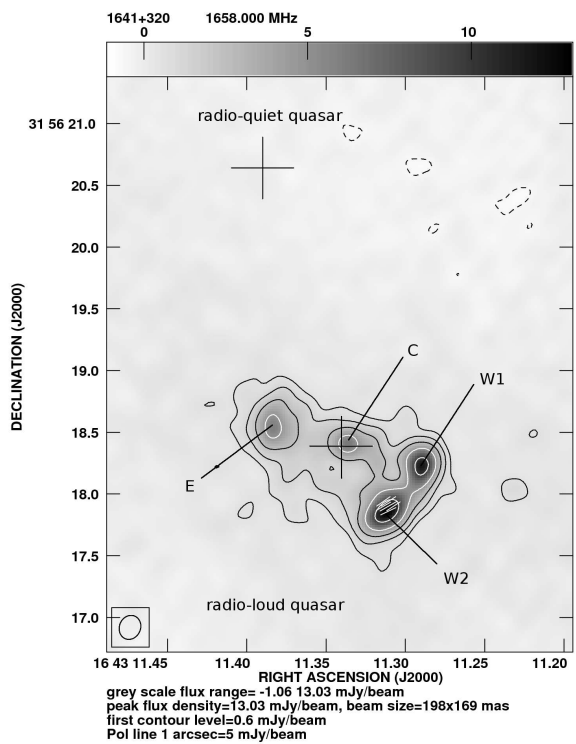}}
\caption{MERLIN radio image at 1.66 GHz of the binary
quasar FIRST 1641+320 \cite{Ref8}. The pair consists of the radio-loud and
radio-quiet sources. Their positions found in the SDSS are marked by crosses.
Contours increase by a factor of two and the first contour level
corresponds to $\sim 3\sigma$. The symbols indicate: C - radio core,
E - eastern jet, W1,W2 - western jet}
\label{fig:2}       
\end{figure}

\section{Mechanisms of the disk stabilization}

The radiation pressure on the theoretical grounds has been verified by
 numerical tests and simulations (e.g., \cite{Ref12}).
Two-dimensional simulations show that the radiation pressure
in fact contributes to the total pressure in the viscous stress tensor,
so such disk may be thermally and viscously unstable.
On the other hand, the exact prescription for the disk heating law, either 
with a sum, or a geometrical mean of the gas and radiation pressures,
is not certain. The former seems somewhat more natural and is elegant to use this simple form due to the Occam's razor. However, it has been shown that 
such a prescription leads in time dependent simulations to very strong 
periodic outbursts of the disk. Such outbursts by many orders of magnitude 
in luminosity are hardly observed, especially in AGN case.

One of the plausible mechanisms discussed in the literature that may 
affect the stability of accretion disks is the jet outflow. 
It has been shown that it correlates with the mass of an accreting black hole,
giving an additional degree of freedom in the so-called 'fundamental plane' 
 \cite{Ref11}. 
On this plane, the X-ray luminosity of a source correlates with its 
radio loudness which is the measure of the jet power. For AGN it may therefore 
result in stronger stabilizing effect on the radiation pressure dominated disks, even if the heating prescription is uniform for any mass of black hole.
This supports the results of the simulations of accretion disk instabilities, 
where the jet power is incorporated in the model and provides additional 
source of cooling to the plasma (\cite{Ref6}).

Another effect that has been found recently (\cite{Ref8}) is a potential 
stabilizing effect of the companion star in a binary system 
or companion galaxy in case of quasars.
The activity modulation due to the changes of the outer boundary 
conditions may be due to the periodically changing external accretion rate, 
e.g. if the binary system orbit is eccentric. In this case the activity cycles 
are separated by prolonged low states, where the accretion rate in the disk 
drops below the level which is sufficient to trigger the instability.
In addition, we found that if a variable energy transport to the jet is 
allowed, a prolonged high state appears between the limit cycle oscillations.
This is an interesting discovery that could explain the effect of 
'superoutbursts', observed in some sources.

Finally, what we found recently is that the viscosity fluctuations, modeled 
stochastically as a Markov chain process (\cite{Ref9}, \cite{Ref10}) will lead to the ultimate viscous 
stability of radiation pressure dominated disk. This occurs
 when the fluctuating part in the viscosity coefficient is rather large.

\begin{figure}
\resizebox{0.95\columnwidth}{!}{
 \includegraphics{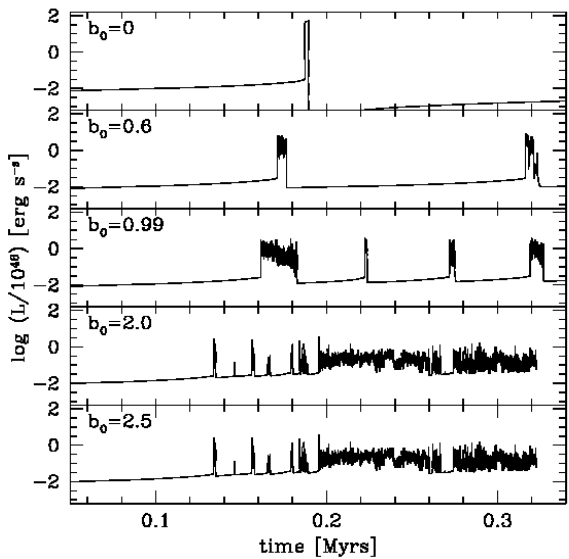}}
\caption{Theoretical lightcurves of the accretion disk in AGN under the limit 
cycle oscillations due to the radiation pressure instability (parameters of 
the model: black hole mass $M= 3 \times 10^{8} M_{\odot}$, accretion rate 
$\dot M = 0.33 \dot M_{\rm Edd}$ and $\alpha_{0}=0.1$.
The 
oscillations are gradually suppressed with the increase of the fluctuating part 
in the viscosity coefficient, as parametrized by the factor $b_{0}$.}
\label{fig:3}       
\end{figure}

In Figure \ref{fig:3} we plot the example of the limit cycle oscillations in 
the AGN accretion disk luminosity due to the radiation pressure. These 
outbursts are gradually stabilized by the fluctuating part 
in the viscosity coefficient, equal to 
$\alpha = \alpha_{0} (1+ b_{0} u_{\rm n} (r,t))$, where $u_{\rm n}$ is changing
according to a Markov chain model, i.e. recalls the value of the $u_{\rm n-1}$. The coherent length of these fluctuations is defined by the disk vertical thickness
while their time is governed by the viscous timescale.
During the activity cycles, a hot disk is 
geometrically thicker than a cold one, so the net effect requires 
numerical modeling and global, time-dependent simulations. Still, the 
thermal fluctuations in such models remain and the resulting lightcurves of
accreting sources exhibit stochastic variability on very short timescales. The 
power density spectra approximately scale inversely with frequency and may be 
compared with observed variable sources, 
if the lightcurves are appropriately long and time-resolved.

\section{Summary}

In accretion disks we can have two main types of thermal-viscous instabilities:
\begin{itemize}
\item Radiation pressure instability
\item Partial hydrogen ionization instability
\end{itemize}
The radiation pressure instability can
 lead to the short term limit cycle oscillations in black hole x-ray
binaries (tens-hundreds seconds scales) or to cyclic activity of quasars 
(scales of tens-thousands of years). The ionization (or dwarf-nova type) 
instability can lead 
to the X-ray novae eruptions (scales of months-years) or 
long-term activity cycles in AGN (scales of millions of
years).

The disk can be stabilized by:
\begin{itemize}
\item Very strong jet/outflow
\item Heating prescription
\item Companion
\item Viscous fluctuations
\end{itemize}

\end{document}